\newcommand{\IA}{{\it IA827}}
\newcommand{\NB}{{\it NB816}}

\newcommand{\Izb}{{\it Iz827}}

\documentstyle[emulateapj]{article}

\begin{document}

\title{A SUBARU SEARCH FOR LYMAN$\alpha$ EMITTERS AT z=5.8
       WITH AN INTERMEDIATE-BAND FILTER}

\author{Masaru Ajiki\altaffilmark{1},
        Yoshiaki Taniguchi\altaffilmark{1},
        Shinobu S. Fujita\altaffilmark{1},
        Yasuhiro Shioya\altaffilmark{1},\\
        Tohru Nagao\altaffilmark{1}\altaffilmark{2},
        Takashi Murayama\altaffilmark{1},
        Sanae F. Yamada\altaffilmark{1},
        Kazuyoshi Umeda\altaffilmark{1},\\
        Shunji S. Sasaki\altaffilmark{1},
        Ryoko Sumiya\altaffilmark{1}, and
        Yutaka Komiyama\altaffilmark{3}
}

\altaffiltext{1}{Astronomical Institute, Graduate School of Science, Tohoku University, \\
                 Aramaki, Aoba, Sendai 980-8578}
\altaffiltext{2}{INAF -- Osservatorio Astrofisico di Arcetri,
                 Largo Enrico Fermi 5, 50125 Firenze, Italy}
\altaffiltext{3}{Subaru Telescope, National Astronomical Observatory of Japan,\\
                 650 North A'ohoku Place, Hilo, HI 96720, USA}



\begin{abstract}
We present the results of a survey for Ly$\alpha$ emitters at
$z \approx$ 5.8 using a new intermediate-band filter
centered at $\lambda_{\rm c} \approx $ 8275 \AA ~ 
with $\Delta\lambda_{\rm FWHM} \approx $ 340 \AA ~
(i.e., the spectroscopic resolution is $R \approx 23$) with
a combination with a traditional narrow-band centered at
$\lambda_{\rm c} \approx$ 8150 \AA ~ with 
 $\Delta\lambda_{\rm FWHM} \approx $ 120 \AA ~
($R \approx 68$). Our observations were made with use of the Subaru 
Prime Focus Camera, Suprime-Cam, on the 8.2 m Subaru telescope
in a sky area surrounding the high redshift quasar,
SDSSp J104433.04$-$012522.2 at $z=5.74$, covering an effective
sky area with $\approx 720$ arcmin$^2$. In this survey,
we have found four Ly$\alpha$-emitter candidates from the intermediate-band image
($z \approx$ 5.8 with $\Delta z \approx 0.3$).
Combined with our previous results based on the \NB~ imaging,
we discuss the star formation activity in galaxies
between $z \approx 5.7$ and $z \approx 5.9$.

\end{abstract}

\section{INTRODUCTION}

Probing the star formation activity in galactic or subgalactic systems
at high redshift is essentially important for understanding of 
the formation and early evolution of galaxies. This is also important 
to investigate major sources of the cosmic reionization (e.g., Loeb \&
Barkana 2001).
Recent optical deep surveys and their optical spectroscopic follow-up
observations have revealed more than two dozens of such 
star forming galaxies beyond $z=5$ 
(e.g., Hu et al. 2002, 2004; Kodaira et al. 2003; Santos et al. 2003;
see for a review Taniguchi et al. 2003b, and Spinrad 2003).

In an attempt to find star-forming objects at $z \approx 5.7$,
we made an optical deep imaging survey 
in a sky area around the SDSS quasar, SDSSp J104433.04$-$012502.2
(Ajiki et al. 2003). In this survey, we found  
20 Ly$\alpha$ emitter (LAE) candidates at $z \approx 5.7$.
Two of them were confirmed as real
LAEs at $z =5.69$ (Ajiki et al. 2002) and at $z=5.66$ 
(Taniguchi et al. 2003a) based on their optical spectroscopy.
In this survey, the narrowband filter \NB~
($\lambda_{\rm c}$ = 8150 \AA ~ with $\Delta\lambda_{\rm FWHM}$ 
= 120 \AA) was used to select LAEs at $z \approx 5.7$.
In addition to this narrowband filter, we also used 
a new  intermediate-band filter \IA~
($\lambda_{\rm c}$ = 8275 \AA ~ with $\Delta\lambda_{\rm FWHM}$
 = 340 \AA) (Hayashino et al. 2001; Taniguchi 2001; 
Taniguchi et al. 2003b) in the same observing run.
Using the \IA~ filter, we are able to search for LAEs
in the very luminous part of
the Ly$\alpha$ luminosity function at $z\sim5.8$
in a very large volume.
This intermediate-band filter is one of a series of 
intermediate-band filters, called the IA filter system,
dedicated to the Suprime-Cam on the 8.2m Subaru telescope 
(e.g., Taniguchi 2001; see also Fujita et al. 2003).
The scientific merits of the IA filter system are described in
Taniguchi (2001), see for some spectacular results with a similar
 composite filter system, COMBO-17, Wolf et al. (2003).

In this paper, we report new results on our deep imaging
survey for LAEs at $z \approx 5.8$ with use of the intermediate-band
filter, \IA.
We adopt a flat universe with $\Omega_{\rm matter} = 0.3$,
$\Omega_{\Lambda} = 0.7$, and $h_{70}=1$ where $h_{70} =
H_0/($70 km s$^{-1}$ Mpc$^{-1}$).
Magnitudes are given in the AB system throughout this paper.

\section{OBSERVATIONS AND DATA REDUCTION}

\subsection{Observations}

We have carried out a very deep optical imaging survey for 
faint LAEs in the field surrounding
the quasar SDSSp J104433.04$-$012502.2 at redshift 5.74\footnote{
The discovery redshift was $z=5.8$ (Fan et al. 2000).
Since, however, the subsequent optical spectroscopic observations
suggested a bit lower redshift; $z=5.73$ (Djorgovski et al. 2001)
and $z=5.745$ (Goodrich et al. 2001), we adopt $z=5.74$ in this paper.}
(Fan et al. 2000; Djorgovski et al. 2001; Goodrich et al. 2001),
using the prime-focus wide-field camera, Suprime-Cam (Miyazaki et
al. 2002) on the 8.2 m Subaru Telescope (Kaifu et al. 2000) 
on Mauna Kea. Suprime-Cam consists of ten 2k$\times$ 4k CCD chips
and provides a very wide field of view,
$34^\prime \times 27^\prime$ (0.202 arcsec pixel$^{-1}$).

\begin{figure}
\epsscale{0.45}
\plotone{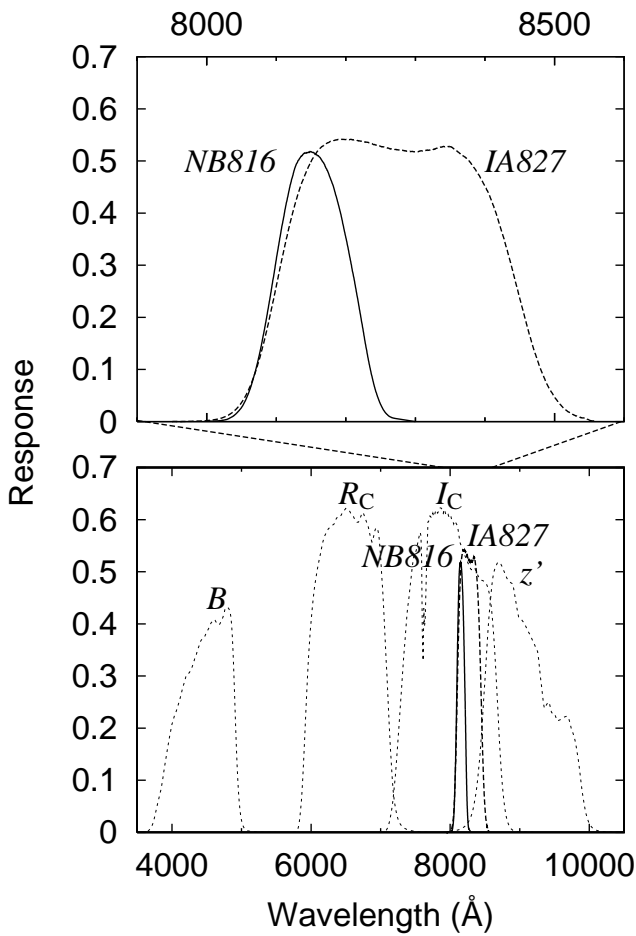}
\caption{Response curves (filter, optics, atmosphere transmission,
         and CCD sensitivity are taken into account)
         of the filters used in our observations. Upper panel shows
         the response curves of both the \NB~ and \IA~ filters.
\label{fil}}
\end{figure}

In this survey, we used the intermediate-band filter,
\IA, centered at 8275 \AA ~
with a passband of $\Delta\lambda_{\rm FWHM} = 340$ \AA; the
wavelength coverage corresponds to the redshift range of 5.66 -- 5.94 for
Ly$\alpha$ emission; note that the central wavelength of 
\IA~ filter varies within 25 \AA~
with the positions on the filter, causing an uncertainty 
in the redshift estimate within $\Delta z \simeq 0.02$.
We also used broad-band filters,
$B$, $R_{\rm C}$, $I_{\rm C}$, and $z^\prime$,
and the narrow-band filter,
\NB, centered at 8150 \AA ~ with a passband of
$\Delta\lambda_{\rm FWHM} = 120$ \AA; the wavelength corresponds
to the redshift range of 5.65--5.75 for Ly$\alpha$ emission (see Ajiki et al. 2003).
The total-response (filter, optics, and atmosphere transmission and 
 CCD sensitivity are taken into account) curves of the filter bands
used in our observations are shown in Figure \ref{fil}.
A summary of the imaging observations is given in Table \ref{tab:obs}.
All observations were done under photometric conditions,
and the seeing was between 0$\farcs$7 and 1$\farcs$4
during the observing run.
Spectrophotometric standard stars used in the flux
calibration for \IA~ are HZ 21 (Oke 1990), and PG 1034+001 (Massey et al. 1988).
The detailed flux calibration for
 the $B$, $R_{\rm C}$, $I_{\rm C}$, and \NB~ data and 
 data reduction procedures are given in Ajiki et al. (2003).

The total size of the reduced field is $32.^\prime08 
\times 25.^\prime23 \approx 807$arcmin$^2$.
After masking the regions contaminated by fringes and bright stars,
our actual survey area is  $\approx$ 720 arcmin$^{2}$ (see Ajiki et al. 2003).
The volume probed by the \IA~ imaging has (co-moving) transverse
 dimensions of 4.0 $\times 10^{3} h_{0.7}^{-2}$ Mpc$^2$, and
 the FWHM of the filter corresponds to a co-moving depth along
the line of sight of 123 $h_{0.7}^{-1}$ Mpc ($z_{\rm min} \approx 5.66$
 and $z_{\rm max} \approx 5.94$; note that these values are average of 13
 different position on the \IA~ filter).
Therefore, a total volume of $5.0 \times 10^{5} h_{0.7}^{-3}$ Mpc$^{3}$
 is probed in our \IA~ image.

\subsection{Source Detection and Photometry}

Source detection and photometry were performed using
 SExtractor version 2.2.2 (Bertin \& Arnouts 1996) in the double image mode.
A source is selected as a 13-pixel connection above the 2 $\sigma$ noise level
 on the \IA~ image.
Photometry was performed with a 2.8 arcsec diameter aperture for each band
 image after matching image size of the data of each band with
 the $R_{\rm C}$-band data (1.4 arcsec).
The limiting magnitudes are $\IA=25.6$, $\NB=26.0$,
 $B=26.6$, $R_{\rm C}=26.2$, $I_{\rm C}=25.9$, and
 $z^\prime=25.3$ for a 3$\sigma$ detection with a 2.8 arcsec diameter
 aperture.
In the above source detection, we find $\sim 34,000$ sources down to
 $\IA = 25.6$.

\section{RESULTS}

\subsection{Selection of \IA-Excess Objects}

Since the effective wavelength of the \IA~ filter is
8275 \AA, we use a continuum band \Izb~ evaluated from
a linear combination of $f_{\Izb}= 0.64 f_{I_{\rm C}} + 0.36
 f_{z^\prime}$ where $f_{I_{\rm C}}$ and $f_{z^\prime}$ are
the fluxes at $I_{\rm C}$ and $ z^\prime$ bands, respectively.
A 3 $\sigma$ limit of
\Izb~ is $\simeq 26.0$ in a 2.8 arcsec diameter aperture.

\begin{figure}
\epsscale{0.45}
\plotone{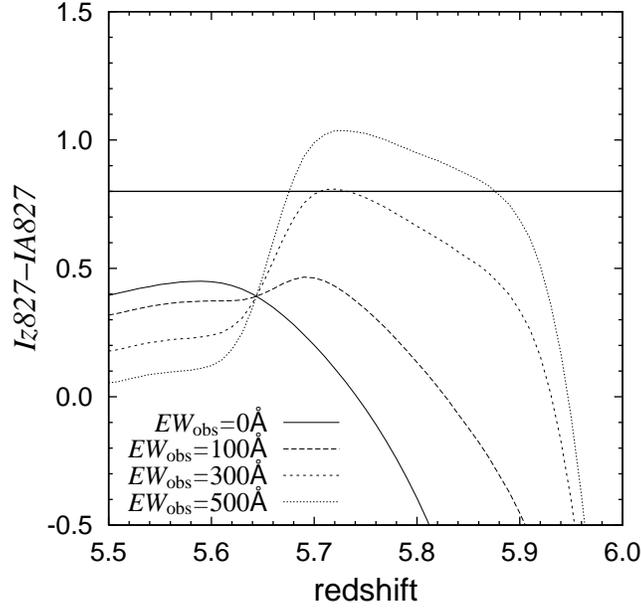}
\caption{The \IA ~ excess of model LAEs with $EW_{\rm obs}=0$ \AA, 100 \AA, 300 \AA,
         and 500 \AA~ as a function of redshift.
         The horizontal line shows our selection criterion for \IA-excess objects.
\label{izia}}
\end{figure}

When we select \IA-excess objects, we should be careful because 
objects at $z\sim5.6$ with no emission line could be selected
as nominal \IA-excess objects. This is due to that 
the absorption by intergalactic neutral hydrogen causes 
the strong continuum depression at $\lambda \sim 8000$ \AA ~
for objects at $z\sim5.6$. In order to exclude such contamination
in our selection procedure of \IA-excess objects, 
we investigate the effect of continuum depression for high-$z$
LAEs with different Ly$\alpha$ emission-line equivalent widths
($EW$ = 0, 100, 300, and 500 \AA)
as a function of redshift; $5.0 \leq z \leq 6.0$). 
The results are shown in Figure \ref{izia}.
In these estimates, we use the average optical depth derived by 
Madau et al. (1996).
Figure \ref{izia} shows that 
even objects at $z\sim5.6$ with no emission line
have \IA~ excesses as much as $\Izb-\IA\sim 0.5$ mag.
We also note that  the redshift range of detectable LAEs depends on EW
(see section \ref{sec:nlae}).

Taking our results shown in Figure \ref{izia},
we adopt a selection criterion for \IA-excess objects, $ \Izb - \IA > 0.8$,
because the error of $\Izb - \IA$ is $\simeq 0.3$ mag for objects
with $\IA=24.5$.
Then, we select \IA-excess objects by using the following criteria;

\begin{eqnarray}
\IA & < & 24.9,\\
\Izb - \IA & > & 0.8,\\
\Izb - \IA & > & 3 \sigma( \Izb - \IA),
\end{eqnarray}
where
\begin{equation}
3 \sigma( \Izb - \IA)=
 -2.5 \log{\left(1-\frac{\sqrt{(3 \sigma_{\IA})^2
 +(3 \sigma_{\Izb})^2}}{f_{\IA}}\right)}.
\end{equation}
The criterion, $3 \sigma(\Izb - \IA)$ corresponds to 
a line flux, $F_{\rm L} \approx 1.5 \times 10^{-17}$ergs s$^{-1}$ cm$^{-2}$.
This line-flux limit is higher by a factor of $\sim 3$ than that in Ajiki et al.(2003).

\begin{figure}
\epsscale{0.45}
\plotone{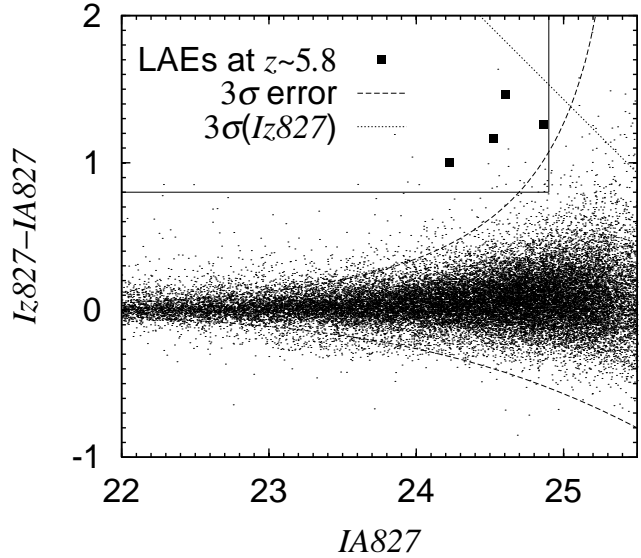}
\caption{Color-maginitude diagram between $\Izb - \IA$ and $\IA$.
         All objects detected down to the apparent magnitude of $\IA=25.6$
         in the \IA-selected catalog are shown.
         The horizontal solid line corresponds to the color of
         $\Izb-\IA=0.8$ and the vertical solid line corresponds to 
         the magnitude limit of $\IA=24.9$.
         Solid curves show the distribution of $3 \sigma$ error.
\label{iacm}}
\end{figure}

In Figure \ref{iacm}, we show the diagram between $\Izb-\IA$ and
\IA~ for the objects in the \IA-selected catalog together with
the above criteria.
There are 21 \IA-selected sources which satisfy the above three criteria.
Note that five of the 21 objects have been also selected as
\NB-excess objects in our previous work (Ajiki et al. 2003).

\subsection{Selection of LAE Candidates}
\label{select}

\begin{figure}
\epsscale{0.45}
\plotone{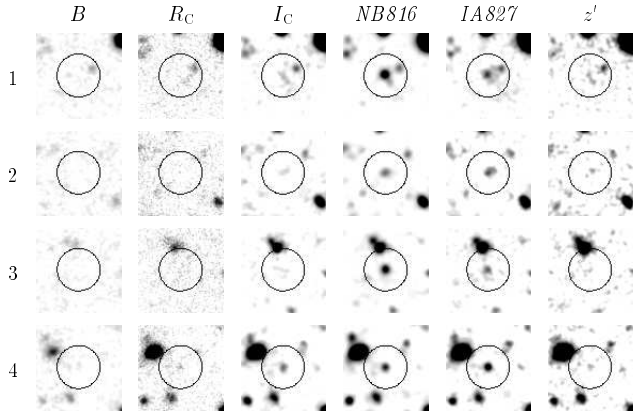}
\caption{$B$, $R_{\rm C}$, $I_{\rm C}$, \NB, \IA, and $z^\prime$ images of
         our LAE candidates at $z \approx 5.8$.
         Each box is $16^{\prime \prime}$ on a side.
         Each circle is $4^{\prime \prime}$ radius.
         The numbers shown in the left column correspond
         to those given in the first column of Table 2.
\label{thum}}
\end{figure}

In order to select LAE candidates at $z \approx 5.8$ from
 our emission-line objects, we apply the same criteria as those in Ajiki et al. (2003)
 to all emitters; i.e.,
\begin{eqnarray}
                 B  & > & 26.6, \label{eq:c1}\\
R_{\rm C}-I_{\rm C} & > & 1.8~~~   {\rm for} ~~ I_{\rm C}  \leq  24.8, \label{eq:c2}\\
          R_{\rm C} & > & 26.6 ~~  {\rm for} ~~ I_{\rm C}  >     24.8.\label{eq:c3}
\end{eqnarray}
These criteria enable us to select LAEs at $z \approx 5.8$
(see Ajiki et al. 2003).
First, eleven of the 21 objects satisfy the criterion (\ref{eq:c1}).
None of the eleven objects has $I_{\rm C}$ magnitude of  $\leq  24.8$.
Therefore no object satisfies the criterion (\ref{eq:c2}).
Four of the eleven object satisfy the criterion (\ref{eq:c3}).
Finally we select 4 objects as LAE candidates at $z \approx 5.8$
by the criteria (\ref{eq:c1}) and (\ref{eq:c3}).
Three of the 4 LAE candidates have been already selected as LAE candidates by 
Ajiki et al. (2003) based on their \NB~ data.
The positions and photometric properties of the four $IA827$-selected
LAEs are given in Table \ref{tab:LAE}.
It is noted that all of our LAE candidates are undetected above 2$\sigma$
level in the $B$- and $R_{\rm C}$-band images
 (i.e., $B>27.0$ and $R_{\rm C}>26.6$).
The $B$, $R_{\rm C}$, $I_{\rm C}$, \NB, \IA,
 and $z^\prime$ images of the four LAE candidates are
shown in Figure \ref{thum}.
The comparison of results of this survey and those of using \NB ~ (Ajiki et al. 2003)
are summarized in Table \ref{tab:nbia}.

\section{DISCUSSION}

\subsection{Properties of the LAE Candidates Expected from the \IA~ and \NB~ Data}

\begin{figure}
\epsscale{0.55}
\plotone{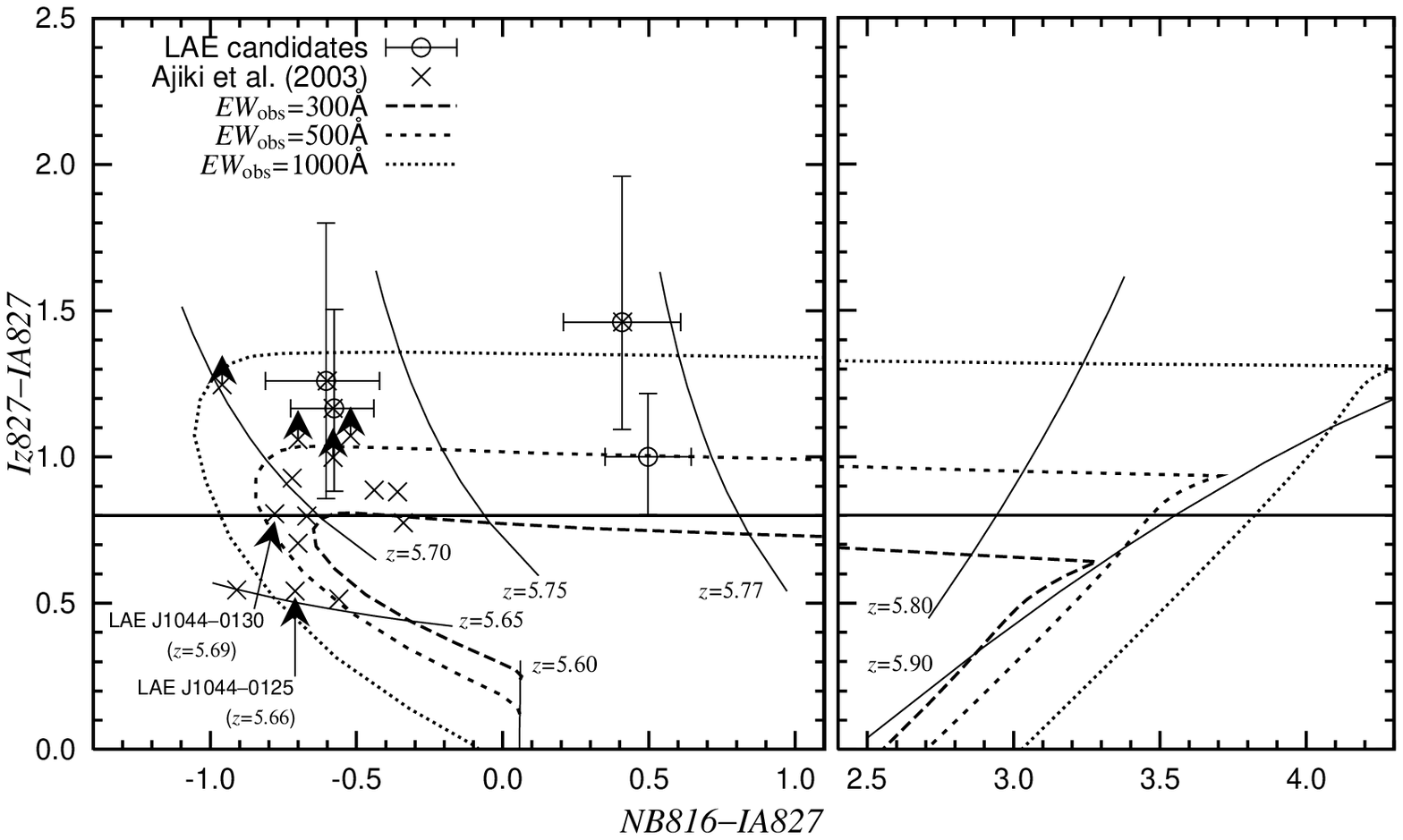}
\caption{Diagram between $\Izb-\IA$ and $\NB-\IA$.
         Four $\IA$-selected LAEs in this study and $\NB$-selected
         LAEs in Ajiki et al. (2003) are shown. 
         Color properties of model LAEs with
         $EW_{\rm obs}=300$ \AA, 500 \AA, and 1000 \AA~ are also shown.
         The horizontal line shows our selection criterion for the \IA-excess objects.
         The two LAEs confirmed by our spectroscopy
         are marked (Ajiki et al. 2002; Taniguchi et al. 2003a).
         Note that none of LAE candidates found by this survey
         has $\NB-\IA>1.0$.
\label{nbia}}
\end{figure}

In the previous section, we selected the 4 LAE candidates at $z\approx5.8$
given in Table \ref{tab:LAE}.
All of them were also detected in our \NB~ image.
We try to estimate redshifts of the 4 LAE candidates using both 
the \NB~ and \IA~ data.
In Figure \ref{nbia}, we show the diagram of 4 LAE candidates at $z\approx 5.8$
between $\Izb-\IA$ and $\NB-\IA$. In this figure, we also show colors
of model LAEs with $EW_{\rm obs}=300$ \AA, 500 \AA, and 1000 \AA.
In these models, we use the average optical depth derived by Madau et al. (1996)
 to estimate the absorption by intergalactic neutral hydrogen,
It is found that the color of the LAE candidates
at $z \approx 5.7$ above the 3 sigma detection in \IA~(Ajiki et al. 2003)
are consistent with model LAEs at $z=$ 5.65 -- 5.75.
It is also found that
two of our 4 LAE candidates are expected to be at $z=$ 5.70 -- 5.75,
and the other two are at $z=$ 5.75 -- 5.77.
We can also estimate the EWs and line fluxes of our 4 LAE candidates
from Figure \ref{nbia}.
The estimated EWs, $EW_{\rm est}$, and line fluxes, $F_{\rm L, est}$,
of our LAE candidates are given in Table \ref{tab:est}.

\subsection{Space Density of the LAEs at $z \approx 5.8$}

\label{sec:nlae}

Since we have detected the 4 LAE candidates at $z \approx 5.8$
in the volume of $5.0 \times 10^{5} h_{0.7}^{-3}$ Mpc$^{3}$,
we obtain the space density of the LAE candidates at $z \approx 5.8$,
$n({\rm Ly}\alpha) \simeq 8 \times 10^{-6} h_{0.7}^{3}$Mpc$^{-3}$.
This density is lower by an order of magnitude than that of our \NB~ survey,
$n({\rm Ly}\alpha) \simeq 1.1 \times 10^{-4} h_{0.7}^{3}$Mpc$^{-3}$ at
$z \approx 5.7$ (Ajiki et al. 2003), or that of Rhoads \& Malhotra (2001),
$n({\rm Ly}\alpha) \simeq 1.0 \times 10^{-4}$ Mpc$^{-3}$ at $z \approx 5.70$
and $n({\rm Ly}\alpha) \simeq 6.4 \times 10^{-5}$ Mpc$^{-3}$ at
$z \approx 5.77$ (see also Rhoads et al. 2003).

The main reason for this seems to be attributed to
different limits in the EW and Ly$\alpha$ luminosity among the above surveys.
The limits in the EW and Ly$\alpha$ luminosity of our survey,
 $EW_{\rm obs, lim}\approx 300$~ \AA~ and
 $L_{\rm Ly\alpha, lim} \approx 1.5 \times 10^{43}$ ergs s$^{-1}$,
 are much higher than those of the other surveys,
 $EW_{\rm obs, lim}\approx 180$~ \AA~ and
 $L_{\rm Ly\alpha, lim} \approx 5.0 \times 10^{42}$ ergs s$^{-1}$
 for that of Ajiki et al. (2003), or
 $EW_{\rm obs, lim}\approx 75$~ \AA~ and
 $L_{\rm Ly\alpha, lim} \approx 2.5 \times 10^{42}$ ergs s$^{-1}$
 for that of Rhoads \& Malhotra (2001).
Therefore, only a few objects that are selected as LAE candidates in the other surveys
satisfy the limits in the EW and Ly$\alpha$ luminosity  of our survey.
Actually, only four of 20 objects found in Ajiki et al. (2003) satisfy these criteria.

\begin{figure}
\epsscale{0.35}
\plotone{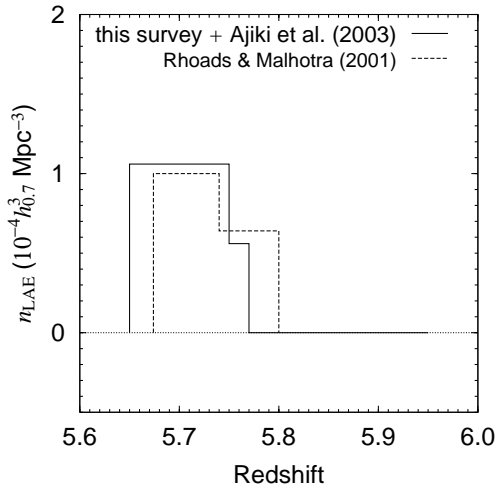}
\caption{Number densities of LAE candidates found both in this survey
         and in Ajiki et al. (2003) as a
         function of redshift are compared with those derived by
         Rhoads \& Malhotra (2001).
         Note that there is no LAE candidate at $z=$5.77--5.94
         in our survey.
\label{nlae}}
\end{figure}

In Figure \ref{nlae}, the space density of both our LAE candidates
and those found in Ajiki et al. (2003)  are shown 
 as a function of redshift together with that of Rhoads \& Malhotra (2001).
The redshift range for our LAE candidates
is classified into the two redshift intervals, $z=$ 5.65 -- 5.75 and $z=5.75$ -- 5.77.
In the redshift estimate of our LAE candidates, we use Figure \ref{nbia}.
Note that LAE candidates undetected in \IA~ have redshifts between $z=5.65$ and $z= 5.75$.
It is also noted that any corrections for the detection completeness is not made
for all the samples.

Although our  \IA~ filter could prove Ly$\alpha$ emitters 
at $z \approx $5.7 -- 5.9, all our 4 LAE candidates lie
at $z \approx $5.7 -- 5.8. Therefore, it is interesting to consider 
why no LAE candidates are found at $z \approx$ 5.8-- 5.9.
For LAEs at $z\gtrsim 5.8$, 
both the Ly$\alpha$ emission and the very weak continuum depressed  at
wavelengths shortward of the Ly$\alpha$ peak are recorded in the  \IA~ image.
Therefore, such objects are not selected as strong  \IA-excess objects.
This effect is indeed found in our simulations shown in Figure \ref{izia}.
The \IA-excess variation in redshift are shown in Figure \ref{izia}.

However, Figure \ref{izia} also shows that 
if LAE candidates with
very large EWs such as No. 2 in Table \ref{tab:est}~
 are present at $z=$ 5.8 -- 5.9,
they could be detected as \IA-excess objects.
Therefore, we cay say that such very bright LAEs at $z=$ 5.8 -- 5.9 do
not exist at least in our survey field.

\subsection{Ly$\alpha$ Luminosity Distributions at $z=$ 5.7 -- 5.8}

\begin{figure}
\epsscale{0.45}
\plotone{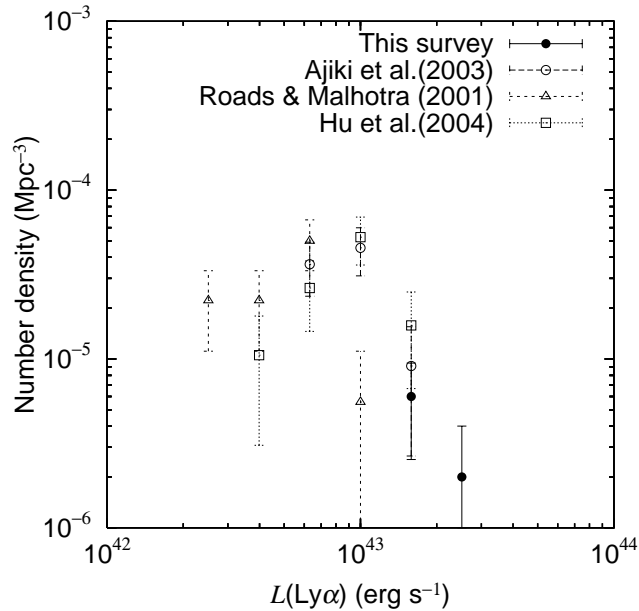}
\caption{Number distributions of LAE candidates found both in this survey
         as a function of Ly$\alpha$ luminosity are compared with those
         derived by Ajiki et al. (2003), Rhoads \& Malhotra (2001),
         and Hu et al. (2004).
         The Ly$\alpha$ luminosity distribution of these surveys are estimated from
         their photmetric catalogs.
\label{lf}}
\end{figure}

We investigate the Ly$\alpha$ luminosities of the $z \approx 5.8$ 
candidates. 
The derived  Ly$\alpha$ luminosities of this survey range from 
$1.8 \times 10^{43}$ to $2.9 \times 10^{43}$ ergs s$^{-1}$.
In Figure \ref{lf}, we show the number distributions of our LAE candidates
 as a function of Ly$\alpha$ luminosity together with those
 previous LAE surveys at $z\approx5.7$ using a narrowband filter
(Ajiki et al. 2003; Rhoads \& Malhotra 2001; Hu et al. 2004).
The figure shows that our survey probes higher-luminosity sources 
with respect to the other LAE surveys at $z\approx5.7$;
 see also Pascarelle et al. (1998) and Fujita et al. (2003).
Since an intermediate-band can cover a wider volume than a typical narrowband filter,
the use of such intermediate-band filter is useful finding higher-luminosity LAEs
 that are rarer than low luminosity ones.

\subsection{Possibility of Large Scale Structure at $z=$ 5.7 -- 5.8}

\begin{figure}
\epsscale{0.45}
\plotone{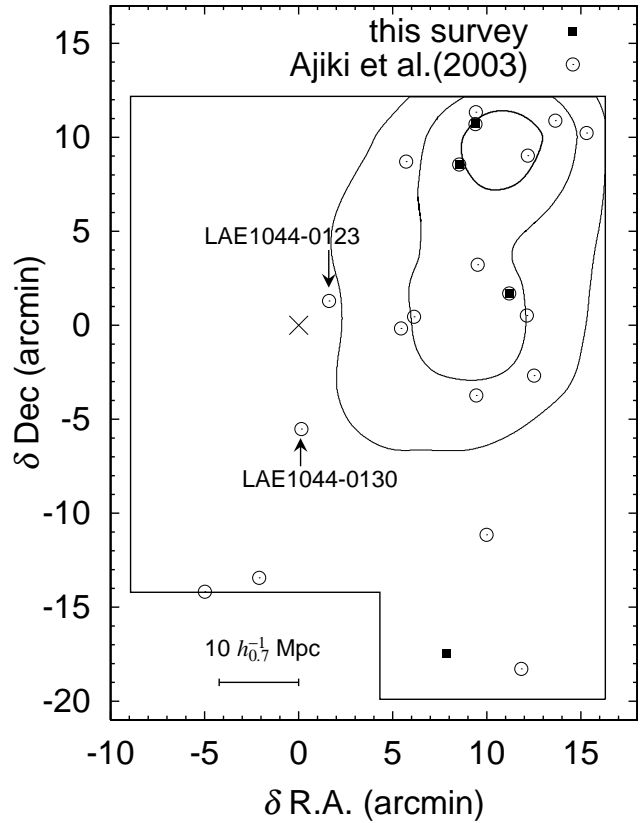}
\caption{Celestial positions of both the 4 LAE candidates
         at $z \approx 5.8$ and those at $z \approx 5.7$ found in Ajiki et al. (2003).
         Our survey area is shown by solid line.
         The position of SDSSp J104433.04$-$012502.2 is shown by
         ``$\times$''.
         The two LAEs at $z\approx 5.7$ confirmed by our spectroscopy
         are marked with open circles (Ajiki et al. 2002; Taniguchi et al. 2003a).
         The contours show the local surface density of the LAE candidates of
         $\overline{\Sigma}$, $2 \overline{\Sigma}$, and $3 \overline{\Sigma}$,
         where $\overline{\Sigma}$ is the surface density averaged over the field,
         $2.9 \times 10^{-2}$ arcmin$^{-2}$.
\label{map}}
\end{figure}

In Figure \ref{map}, we plot the spatial distributions the 4 LAE candidates
 together with those at $z\approx 5.7$ of Ajiki et al. (2003).
It appears that most LAE candidates are found in the western side of
 the quasar SDSSp J104433.04$-$012502.2.
In particular, there is no LAE candidate in the northeast of the quasar.
Contours of the local surface density are also shown in this figure.
The local surface density at position ($x$, $y$) is the density averaged 
over the circle centered at ($x$, $y$) whose radius is determined as
the angular  distance to 5 nearest neighbors. Note that 
smoothing with the top-hat filter of $\approx3.4$ arcmin, corresponding  to
$\approx 8 h^{-1}_{0.7}$ Mpc at $z\sim5.7$, is made.

The contour map suggests that there is a high-density region in northwest side of the quasar,
where the local density is higher by a factor of 3 than the average density
 in this field.
To examine its statistical significance,
we made simple simulations; we distributed 21 points randomly in the survey field
and estimated the local surface density at each point.
We found that 19 \% of 100 random distributions also show a similar
high-density region.
Therefore, it is difficult to conclude that there is a high-density 
clustering region of LAEs at $z\approx5.7$.
 
\vspace{4ex}

We would like to thank both the Subaru and Keck Telescopes staff for
their invaluable help, and Dave Sanders, Sylvain Veilleux, S. Okamura,
Y. Ohyama, S. Oyabu, N. Kashikawa, M. Iye, H. Ando, and H. Karoji
 for encouragement during the course of this study.
We would also thank James Rhoads for providing us useful information on
 the LALA survey, M. Ouchi, M. Yagi, and K. Shimasaku for useful discussion
 on the data reduction of Suprime-Cam data,
 and T. Hayashino for his technical help.
We also thank to the referee, Paul Francis, for his useful comments and
 suggestion.
This work was financially supported in part by
the Ministry of Education, Culture, Sports, Science, and Technology
(Nos. 10044052, and 10304013) and JSPS (No. 15340059). 
MA and TN are JSPS fellows.



\begin{deluxetable}{lcccc}
\tablenum{1}
\tablecaption{Journal of observations\label{tab:obs}}
\tablewidth{0pt}
\tablehead{
\colhead{Band} &
\colhead{Obs. Date (UT)} &
\colhead{$T_{\rm int}$ (sec)\tablenotemark{a}}  &
\colhead{$m_{\rm lim}$(AB)\tablenotemark{b}} &
\colhead{$FWHM_{\rm star}$ (arcsec)\tablenotemark{c}}
}
\startdata
$B$            & 2002 February 17      &  1680 & 26.6 & 1.2 \\
$R_{\rm C}$    & 2002 February 15, 16  &  4800 & 26.2 & 1.4 \\
$I_{\rm C}$    & 2002 February 15, 16  &  3360 & 25.9 & 1.2 \\
$z'$           & 2002 February 15, 16  &  5160 & 25.3 & 1.2 \\
\NB            & 2002 February 15 - 17 & 36000 & 26.0 & 0.9 \\
\IA            & 2002 February 15 , 17 & 12420 & 25.6 & 1.2 \\
\enddata
\tablenotetext{a}{Total integration time.}
\tablenotetext{b}{The limiting magnitude (3$\sigma$) within a
2.8 arcsec aperture.}
\tablenotetext{c}{The full width at half maximum of stellar
objects in the final image}
\end{deluxetable}

\begin{deluxetable}{rcccccccr}
\tablenum{2}
\tablecaption{Photometric properties of the LAE candidates at
 $z \approx 5.8$ \tablenotemark{a} \label{tab:LAE}}
\tablewidth{0pt}
\tablehead{
\colhead{No.} &
\colhead{$\alpha$(J2000)} &
\colhead{$\delta$(J2000)} &
\colhead{$I_{\rm C}$\tablenotemark{b}} &
\colhead{\NB\tablenotemark{b}} &
\colhead{\IA\tablenotemark{b}} &
\colhead{$z'$\tablenotemark{b}} &
\colhead{\Izb\tablenotemark{b}} &
\colhead{A03 \tablenotemark{c}} \\
\colhead{} &
\colhead{h ~~ m ~~ s} &
\colhead{$^\circ$ ~~ $^\prime$ ~~ $^{\prime\prime}$}  &
\colhead{}&
\colhead{}&
\colhead{}&
\colhead{}&
\colhead{}&
\colhead{}
}
\startdata
 1 & 10 43 48.3 & $-$01 23 20 &   25.7 & 23.9 & 24.5 &  (25.5) &  25.7  &  7 \\
 2 & 10 43 55.5 & $-$01 14 18 & (26.2) & 25.0 & 24.6 & $>$25.7 & (26.2) & 12 \\
 3 & 10 43 59.0 & $-$01 16 27 & (26.0) & 24.3 & 24.9 & $>$25.7 & (26.2) & 13 \\
 4 & 10 44 01.6 & $-$01 42 31 &   25.0 & 24.7 & 24.2 & $>$25.7 &  25.2  & \nodata \\
\enddata
\tablenotetext{a}{All of our LAE candidates are undetected above 2$\sigma$ level
                 in $B$-and $R_{\rm C}$-band images
                 ($B>27.0$ and $R_{\rm C}>26.6$).}
\tablenotetext{b}{AB magnitude in a 2.8 arcsec diameter.
                  The magnitudes between the 2 $\sigma$ and 3 $\sigma$ detection
                  levels are put in parentheses.}
\tablenotetext{c}{ID number in Ajiki et al. (2003).}
\end{deluxetable}

\begin{deluxetable}{ccccccc}
\tablenum{3}
\tablecaption{Comparison of LAE surveys in SDSSp J1044-0125 field
 \label{tab:nbia}}
\tablewidth{0pt}
\tablehead{
\colhead{Filter\tablenotemark{1}} &
\colhead{$z_{\rm c}$\tablenotemark{2}} &
\colhead{$(z_{\rm max}, z_{\rm min})$\tablenotemark{3}} &
\colhead{$V$\tablenotemark{4}} &
\colhead{$EW_{\rm 0, lim}$\tablenotemark{5}} &
\colhead{$L_{\rm lim}$\tablenotemark{6}} &
\colhead{$N_{\rm LAE}$\tablenotemark{7}} 
}
\startdata
  \NB   & 5.70 & (5.65, 5.75) & 1.8 & 32 &  5 & 20 \\
  \IA   & 5.80 & (5.66, 5.94) & 5.0 & 57 & 15 &  4 \\
\NB~\& \IA \tablenotemark{8}
        & 5.70 & (5.66, 5.75) & 1.7 & 57 & 15 &  3 \\
\enddata
\tablenotetext{1}{Filter name used in the survey.}
\tablenotetext{2}{Central redshift corresponding to the center of the passband.}
\tablenotetext{3}{Minimum and maximum redshift.}
\tablenotetext{4}{Co-moving volume in units of
                  $10^{5} h_{0.7}^{-3}$ Mpc$^{3}$.}
\tablenotetext{5}{Survey limit in rest-frame equivalent width in units of \AA.}
\tablenotetext{6}{Survey limit in Ly$\alpha$-luminosity in units of
                  $10^{42}$ ergs s$^{-1}$.}
\tablenotetext{7}{Number of LAE candidates detected in the survey.}
\tablenotetext{8}{Overlap of the \NB~ and \IA~ surveys}
\end{deluxetable}

\begin{deluxetable}{rccccc}
\tablenum{4}
\tablecaption{ The estimated properties of LAE candidates at $z \approx 5.8$
\label{tab:est}}
\tablewidth{0pt}
\tablehead{
\colhead{No.} &
\colhead{\NB$-$\IA} &
\colhead{\Izb$-$\IA} &
\colhead{$z_{\rm est}$} &
\colhead{$EW_{\rm obs, est}$} &
\colhead{$F_{\rm L, est}$} \\
\colhead{} &
\colhead{} &
\colhead{} &
\colhead{} &
\colhead{(\AA)} &
\colhead{($10^{-17}$ergs s$^{-1}$ cm$^{-2}$)} 
}
\startdata
 1 & $-0.58^{+0.15}_{-0.13}$ & $1.17^{+0.29}_{-0.32}$ & 5.74 & $660^{+590}_{-330}$ & 6.3 \\
 2 & $ 0.41^{+0.21}_{-0.19}$ & $1.46^{+0.44}_{-0.40}$ & 5.77 & $1290^{+4300}_{-640}$ & 6.8 \\
 3 & $-0.60^{+0.19}_{-0.20}$ & $1.26^{+0.48}_{-0.43}$ & 5.74 & $810^{+2800}_{-500}$ & 4.8 \\
 4 & $ 0.50^{+0.16}_{-0.14}$ & $1.00^{+0.17}_{-0.23}$ & 5.77 & $500^{+200}_{-190}$ & 8.0 \\
\enddata
\end{deluxetable}

\end{document}